\documentclass[aip,jcp,reprint,amsmath,amssymb]{revtex4-2}
\usepackage{lmodern}
\usepackage{graphicx}
\usepackage[suffix=]{epstopdf}
\usepackage{natmove}

\newcommand{\vv}[1]{\mathbf{#1}}
\renewcommand{\d}[1]{\ensuremath{\operatorname{d}\!{#1}}}

\begin{document}

\title{Effects of skewing collision cells on transport properties in multiparticle collision dynamics simulations}

\author{Jinny Cha}
\affiliation{Department of Chemical Engineering, Auburn University, Auburn, AL 36849, USA}

\author{Wilfred Kwabena Darko}
\affiliation{Department of Chemical and Biomolecular Engineering, University of Houston, Houston, TX 77204, USA}

\author{Jeremy C. Palmer}
\affiliation{Department of Chemical and Biomolecular Engineering, University of Houston, Houston, TX 77204, USA}

\author{Michael P. Howard}
\email{mphoward@auburn.edu}
\affiliation{Department of Chemical Engineering, Auburn University, Auburn, AL 36849, USA}

\begin{abstract}
Multiparticle collision dynamics (MPCD) is a mesoscale simulation technique that uses a simplified solvent to model hydrodynamic interactions. Rather than interact through pairwise forces, MPCD solvent particles undergo momentum-exchanging collisions within spatially localized cells according to prescribed rules. The conventional MPCD algorithm employs cubic collision cells, but this choice is not optimal for systems that are most naturally described using skewed simulation boxes.  Here, we investigate the behavior of a modified MPCD scheme in which the collision cells are aligned with the vectors that define a triclinic (parallelepiped) simulation box. We find that skewing the collision cells has a small but statistically significant impact on the transport properties of the pure solvent. Similar, but more pronounced, effects are found for nearly hard spheres in solution, including a significant decrease in their nominal self-diffusion coefficient and unphysical anisotropy in their self-diffusion tensor. Thus, our analysis indicates that skewed MPCD collision cells may result in spurious behavior and should be used with caution. We posit that these artifacts may be mitigated by grid-free schemes for placing particles into collision cells. 
\end{abstract}

\maketitle

\section{Introduction}
Solvent-mediated hydrodynamic interactions (HI) transfer momentum between objects dissolved or suspended in fluids and thus strongly influence the dynamics of many soft materials, including colloidal suspensions \cite{ressel:cup:1989, guazzelli:cup:2017}, polymer solutions \cite{kirkwood:jcp:1948, erpenbeck:jcp:1958}, and biological systems \cite{sear:physrev:2019, pierce:physrev:2018}.
Accurate treatment of HI in models for soft materials is hence essential; however, such HI are challenging to describe because the dissolved or suspended objects are often much larger and slower than solvent molecules, creating a significant separation of length and time scales that can prohibit resolving both with full detail \cite{padding:physrev:2006}. To address this challenge, several mesoscale simulation techniques for modeling HI have been developed, including lattice-based approaches such as the lattice-Boltzmann technique \cite{chen:rev:1998, akker:rev:2018}, implicit-solvent methods such as Brownian \cite{allen:book:2017} or Stokesian \cite{brady:rev:1988} dynamics, and particle-based methods such as dissipative particle dynamics  \cite{hoogerbrugge:epl:1992, espanol:jcp:2017} and multiparticle collision dynamics \cite{malevanets:jcp:1999} (MPCD).

MPCD, which is the focus of this article, uses a highly simplified solvent model to produce HI at reduced computational cost \cite{gompper:springer:2009, howard:coche:2019, kapral:jws:2008}. Rather than interact by pairwise forces, the solvent particles periodically undergo collective momentum-exchanging collisions within spatially localized cells according to a prescribed rule. Stochastic rotation dynamics (SRD) is a well-known collision rule for MPCD \cite{malevanets:jcp:1999}, but others have been proposed \cite{allahyarov:physrev:2002, noguchi:epl:2007}. The collisions must conserve mass and linear momentum, but they can be designed to additionally conserve kinetic energy \cite{malevanets:jcp:1999} or angular momentum \cite{noguchi:epl:2007, noguchi:physrev:2008, gotze:physrev:2007, yang:physrev:2015, theers:sm:2014}. A thermostat may also be added to the collisions to produce the correct kinetic temperature \cite{huang:jcp:2010}. Objects can be coupled to the solvent in different ways \cite{malevanets:epl:2000, poblete:physrev:2014, bolintineanu:cpm:2014}, and MPCD has been been widely used to study both equilibrium and nonequilibrium behaviors of a variety of soft materials \cite{gompper:springer:2009,kapral:jws:2008,howard:coche:2019,theers:sm:2014,huang:jcp:2010,huang:macro:2010, nikoubashman:sm:2013,nikoubashman:macro:2017,howard:jcp:2020,chen:mm:2018,chen:sm:2019,wani:jcp:2022,kotkar:aml:2023,wani:sm:2024}.

In the standard MPCD algorithm, the collision cells are defined to be cubes with edge length $\ell$ aligned with the Cartesian axes \cite{malevanets:jcp:1999}, so the simulation box must be, at minimum, orthorhombic with edge lengths that are multiple of $\ell$. Although this choice is simple and robust, it is not optimal for systems that are most naturally modeled using skewed simulation boxes. As one example, a hexagonal array of cylindrical obstacles can be represented with a minimal repeating unit if a monoclinically skewed box is used, whereas orthorhombic boxes require replicating cylinders \cite{nikoubashman:sm:2013}. Another such case occurs in nonequilibrium simulations of uniform shear flow with Lees--Edwards boundary conditions \cite{lees:jpc:1972, todd:book:2017, evans:book:2008}. One way to enforce these boundary conditions efficiently is to continuously deform the simulation box with the flow \cite{evans:mp:1979, evans:ms:1994}. The particles must be remapped into an initially orthorhombic box before each collision if cubic cells are used, whereas no remapping is needed if the cells align with the deformed box. These examples highlight potential advantages of allowing skewed collision cells in the MPCD algorithm; however, it is currently unknown to what extent skew may alter the transport properties of the MPCD solvent or objects embedded in it \cite{kikuchi:jcp:2003, ihle:physrev:2003a, ihle:physrev:2003b}.

In this work, we modified the standard MPCD algorithm to support collision cells aligned with the vectors that define a triclinic (parallelepiped) simulation box and assessed the impact of skew on transport properties of both the pure solvent and solutions of nearly hard spheres. We found that solvent properties were only weakly sensitive to cell skew, but the hard-sphere solutes had reduced diffusion and unphysical anisotropic motion when the collision cells were skewed. These effects persisted across a range of solute concentrations and solvent densities typically used in MPCD simulations. Our findings highlight that while skewed collision cells offer flexibility in MPCD, they should be used with caution.

The rest of this article is organized as follows. Section \ref{sec:methods} describes the modifications made to the MPCD algorithm and provides details of the simulations. Section \ref{sec:results} presents our analysis of the transport properties of both the pure solvent and solutions of nearly hard spheres, including velocity autocorrelation functions, diffusion coefficients, and shear viscosities, with skewed collision cells. Section \ref{sec:conclusions} summarizes our main findings and discusses the implications of using skewed collision cells in MPCD.

\section{Methods}
\label{sec:methods}
\subsection{Algorithm}
\begin{figure}
    \centering
    \includegraphics{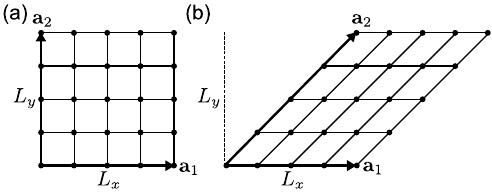}
    \caption{Schematic of collision cells in two dimensions: (a) cubic and (b) skewed cells with $f_{xy} = 1.0$ ($45^\circ$ angle). Both representations describe the same underlying lattice.}
    \label{fig:schematic}
\end{figure}

The MPCD solvent is a fluid of point particles whose positions and velocities evolve discretely in alternating streaming and collision steps \cite{howard:coche:2019, gompper:springer:2009}. During the steaming step, each particle moves according to Newton's equations of motion. The particles are typically force free, but they may be acted on by a body force to simulate flow. During the collision step, the particles are divided into cells and exchange momentum within the cell according to the chosen rule. In the standard MPCD algorithm, these cells are a grid of cubes with edge length $\ell$ aligned to the Cartesian axes [Fig.~\ref{fig:schematic}(a)], but the grid is randomly shifted along each axis by an amount uniformly drawn in $[-\ell/2,\ell/2]$ before every collision to ensure Galilean invariance \cite{ihle:physrev:2001}.

To generalize beyond cubic collision cells, we modified the implementation of MPCD in HOOMD-blue version 4.7.0 to support a grid of collision cells aligned with the three vectors $\vv{a}_1$, $\vv{a}_2$, and $\vv{a}_3$ that define a triclinic (parallelepiped) simulation box,
\begin{equation}
\begin{bmatrix}
\vv{a}_1 & \vv{a}_2 & \vv{a}_3
\end{bmatrix}
=
\begin{bmatrix}
L_x & f_{xy} L_y & f_{xz} L_z \\
0   & L_y        & f_{yz} L_z \\
0   & 0          & L_z
\end{bmatrix},
\label{eqn:box-definition}
\end{equation}
where $L_x$, $L_y$, and $L_z$ are the three edge lengths of an undeformed, orthorhombic box and $f_{xy}$, $f_{xz}$, and $f_{yz}$ are the ``tilt factors'' that skew the vectors relative to each other. We divided the simulation box into an integer number of collision cells along each vector, so the collision cells had the same tilt factors but smaller edge lengths than the simulation box [Fig.~\ref{fig:schematic}(b)]. We shifted the grid randomly in the positive or negative direction along each vector by up to half the spacing between cells. Note that the simulation box has the same volume even when it is skewed, and consequently, the collision-cell volume and the average number of MPCD particles per cell also remain unchanged by skewing. Additionally, if all tilt factors are zero, this formulation reduces to the conventional cubic-cell algorithm. Support for skewed collision cells was released in HOOMD-blue version 5.0.0. We used a mix of both our development version and this publicly released version to perform the simulations in this work.

\subsection{Models}
We simulated both a pure solvent and a solution of nearly hard spheres in simulation boxes with the same volume but differing amounts of skew. The boxes had periodic boundary conditions in all directions. Quantities will be reported in a system of units defined by the mass of a solvent particle $m$, the edge length of a cubic collision cell $\ell$, and the thermal energy $k_{\rm B} T$. The corresponding unit of time is $\tau = \sqrt{m \ell^2/(k_{\rm B} T)}$.

The MPCD solvent was randomly initialized in the simulation box at number density $\rho = 5 / \ell^3$. We employed the SRD collision rule without angular momentum conservation \cite{malevanets:jcp:1999} using a rotation angle of $130^\circ$ about an axis randomly chosen from the unit sphere and a cell-level Maxwell--Boltzmann thermostat \cite{huang:jcp:2010}. The time between collisions was $0.1\,\tau$. The particles streamed between collisions in a single step when they were force free and using a velocity-Verlet integration scheme \cite{allen:book:2017} with timestep $0.01\,\tau$ if a body force was applied.

The solutions of nearly hard spheres used the same solvent model but added solute particles at number density $\rho_{\rm s} = 0.2/\ell^3$. The solute particles had mass $5\,m$ and nominal diameter $1.0\,\ell$. They interacted with each other pairwise through the purely repulsive Weeks--Chandler--Andersen potential,\cite{anderson:jcp:1972}
\begin{equation}
u(r) =
\begin{cases}
4 k_{\rm B} T \left[ \left( \dfrac{\ell}{r} \right)^{12} - \left( \dfrac{\ell}{r} \right)^6 + \dfrac{1}{4} \right], & r \leq 2^{1/6}\ell \\
0, & r > 2^{1/6}\ell
\end{cases},
\label{eqn:wca}
\end{equation}
where $r$ is the particle separation distance. 
The solute particles were coupled to the solvent by including them in the collisions \cite{malevanets:epl:2000}. Between collisions, the solute particles were propagated using conventional molecular dynamics (MD) methods and a velocity-Verlet integration scheme with timestep $0.005\,\tau$. This hybrid MD--MPCD scheme has been widely employed to model solvent-mediated HI in solutions of polymers \cite{huang:macro:2010, nikoubashman:macro:2017} and other complex solutes composed of small spheres \cite{poblete:physrev:2014, wani:jcp:2022, wani:sm:2024, kobayashi:sm:2020}. When used in conjunction with standard cubic MPCD collision cells, it typically gives transport properties in excellent agreement with theoretical expectations.

\subsection{Transport properties}
We characterized the velocity autocorrelation function for the pure solvent as well as the self-diffusion coefficient and shear viscosity for both the pure solvent and the nearly hard spheres in solution (``solutes''). For all quantities, we repeated our measurements a certain number of times, and we report the mean and the uncertainty (one standard error of the mean) across these independent replicas. The simulation protocols for determining each quantity are described next.

\subsubsection{Velocity autocorrelation function}
We first measured the normalized velocity autocorrelation function (VACF) of the pure solvent,
\begin{equation}
c_{v}(t) = \frac{\langle \vv{v}(t) \cdot \vv{v}(0) \rangle}{\langle \vv{v}(0) \cdot \vv{v}(0) \rangle},
\label{eqn:vacf}
\end{equation}
where $\vv{v}$ is the velocity of a particle at time $t$. The solvent was equilibrated for $10^2\,\tau$, then a production simulation of $10^6\,\tau$ was performed during which the VACF was computed every $0.1\,\tau$ up to time $10\,\tau$ on the fly by averaging over the $10^5$ equally spaced independent time origins and all of the solvent particles. This procedure was repeated three times.

\subsubsection{Self-diffusion coefficient}
We next measured the self diffusion of both the pure solvent and solutes through their displacements $\Delta \vv{r}(t)$ after time $t$. The self-diffusion tensor $\vv{D}$ can be extracted from the long-time displacement using,
\begin{equation}
\vv{D} = \frac{1}{2} \lim_{t \to \infty} \frac{\d{}}{\d{t}} \left\langle \Delta \vv{r}(t) \Delta \vv{r}(t) \right\rangle.
\label{eqn:diffusivity-tensor}
\end{equation}
The self-diffusion coefficient $D$ is proportional to the trace of this tensor,
\begin{equation}
D = \frac{1}{3} \mathrm{tr}(\vv{D}),
\label{eqn:diffusivity-trace}
\end{equation}
and can be evaluated directly from the mean squared displacement (MSD) $|\Delta\vv{r}|^2$. Periodic boundary conditions are well-known to introduce finite-size effects on diffusion due to long-ranged HI between images \cite{dunweg:jcp:1993, yeh:jpc:2004}. These effects can be corrected to obtain the self-diffusion coefficient in an infinitely large box $D^\infty$ by evaluating the long-ranged HIs over the repeating lattice of simulation boxes using Ewald summation, \cite{dunweg:jcp:1993, yeh:jpc:2004}
\begin{align}
D^\infty &= D + \frac{k_{\rm B} T}{6 \pi \mu}\Bigg[\sum_{\vv{n} \ne \vv{0}} \frac{{\rm erfc}(\alpha |\vv{n}|)}{|\vv{n}|} \nonumber \\
&+ \sum_{\vv{k} \ne \vv{0}} \frac{4\pi e^{-|\vv{k}|^2/(4\alpha^2)}}{|\vv{k}|^2 V}
-\frac{2\alpha}{\pi^{1/2}} - \frac{\pi}{V\alpha^2} \Bigg],
\label{eqn:diffusivity-ewald}
\end{align}
where $\mu$ is the shear viscosity, the first sum runs over all integer combinations of the box vectors $\vv{n}$ excluding the zero vector, the second sum runs over all reciprocal lattice vectors $\vv{k}$ excluding the zero vector, and $\alpha > 0$ is an arbitrary convergence factor. For a cubic simulation box with edge length $L$, the result of evaluating Eq.~\eqref{eqn:diffusivity-ewald} is well-known to be\cite{hasimoto:jfm:1959, dunweg:jcp:1993, yeh:jpc:2004} 
\begin{equation}
D^\infty = D + \xi\frac{ k_B T}{6 \pi \mu L}
\label{eqn:diffusivity-finite}
\end{equation}
with $\xi \approx 2.837297$.

The pure solvent was equilibrated for $2 \times 10^4\,\tau$, then a production simulation of $10^6\,\tau$ was performed. The MSD was computed every $0.1\,\tau$ up to time $10^2\,\tau$ on the fly by averaging over $10^4$ evenly-spaced time origins and all solvent particles. The self-diffusion coefficient $D$ was calculated by averaging the numerically estimated time derivative of the MSD between $50\,\tau \leq t \leq 100\,\tau$. This procedure was repeated three times.

The solutions of nearly hard spheres were equilibrated for $10^3\,\tau$, then a production simulation of $10^5\,\tau$ was performed during which configurations of the solute were recorded every $10\,\tau$. The solute MSD was calculated up to $10^5\,\tau$ using all configurations as time origins and averaging over all particles. The self-diffusion coefficient was calculated in the same way as for the pure solvent but fit over $10^3\,\tau \leq t \leq 5 \times 10^{3}\,\tau$. The number of replicas depended on the size of the box and is listed in Table S1.

\subsubsection{Shear viscosity}
We last measured the shear viscosity $\mu$ of the pure solvent and solutions using two independent techniques: force-driven simulations \cite{hess:jcp:2002} and reverse nonequilibrium simulations \cite{muller:physrev:1999} (RNES). In the force-driven simulations, we applied a sinusoidal body force,
\begin{equation}
\mathbf{F} = F \sin(k y)\mathbf{e}_x,
\label{eqn:sinusoidal-force}
\end{equation}
to the solvent particles, where $F$ is the force amplitude, $k$ is a wavenumber commensurate with the simulation box, and $\mathbf{e}_x$ is the unit vector in the $x$ direction. This force generates a mass-averaged velocity $u_x$ in only the $x$ direction, which is expected to be
\begin{equation}
u_x(y) = \frac{\rho F}{\mu k^2} \sin(ky)
\label{eqn:velocity_profile}
\end{equation}
for the incompressible flow of a Newtonian fluid. The velocity field measured in the simulations can hence be used to determine the viscosity $\mu$.

We used $F= 0.01171\,k_{\rm B}T/\ell$ and $k = 2\pi/L_y$ for all force-driven simulations so that we had a maximum velocity of roughly $0.15\,\ell/\tau$ for the pure solvent and one periodic repeat in an orthorhombic box. The pure solvent was simulated for $10^{3}\,\tau$ to allow the flow to reach steady state, then a production simulation of $5 \times 10^4\, \tau$ was performed during which the velocity $u_x(y)$ was computed every $1\,\tau$ using a bin width of $0.2\,\ell$ in the $y$ direction. The center-of-mass velocity for each particle configuration was set to zero before averaging to remove small numerical drift. The solutions used a similar protocol with a longer production period of $10^5\,\tau$ and the velocities of all particles sampled every $10\,\tau$. The velocities measured in the simulations were fit to Eq.~\eqref{eqn:velocity_profile} using the viscosity as the fitting parameter. This procedure was repeated ten times for the pure solvent and three times for the solutions.

In the RNES \cite{muller:physrev:1999}, shear flow was generated by an artificial procedure that exchanged the $x$-component of the momentum between solvent particles found in planar slabs with normal oriented along the $y$ direction. The slabs were nonoverlapping and had thickness $\Delta y$ in the $y$ direction. The lower slab consisted of all solvent particles with $-L_y/2 \le y \le -L_y/2+\Delta y$, and the upper slab consisted of all solvent particles with $0 \le y \le \Delta y$. Periodically, a fixed number of pairs of particles having momenta with $x$-component closest to target values of $mV_x$ in the lower slab and $-mV_x$ in the upper slab exchanged the $x$-components of their momenta \cite{tenney:jcp:2010}. After reaching steady state, the cumulative amount of momentum exchanged $p_x(t)$ was recorded as a function of the elapsed time $t$ along with the mass-averaged velocity $u_x(y)$. The shear rate $\dot\gamma=\d{u_x}/\d{y}$ was extracted from the slopes of linear fits of $u_x(y)$ between $-8\,\ell \le y \le -1\,\ell$ and $2\,\ell \le y \le 9\,\ell$, and the average rate of momentum transfer $\dot p_x$ was determined from the slope of a linear fit of $p_x(t)$. The viscosity was then calculated as \cite{muller:physrev:1999}
\begin{equation}
    \mu = \frac{1}{2L_{x}L_{z}} \frac{\dot p_{x}}{\dot\gamma}.
    \label{eq:viscosity}
\end{equation}

Momentum was exchanged between 10 solvent-particle pairs every $0.5 \,\tau$ for the pure solvent and every $0.3\,\tau$ for the solutions using slab thickness $\Delta y = 1\,\ell$. The pure solvent was simulated for $5 \times 10^5 \,\tau$ to reach steady state then simulated for an additional $2.5 \times 10^6 \,\tau$, recording $p_x$ and sampling $u_x$ every $5 \times 10^2\,\tau$; whereas, the solutions were simulated for $2.5 \times 10^5 \,\tau$ to reach steady state then simulated for an additional $5 \times 10^5 \,\tau$ with sampling every $25\,\tau$. The bin width for the velocity sampling was $1\,\ell$ in the $y$ direction. We tested for dependence of $\mu$ on the shear rate by varying the target velocity $V_x$; we used $V_x = \{ 1 , 5 \} \,{\ell}/{\tau}$ for the pure solvent and $V_x = \{ 0.1, 0.5, 1 , 5 \} \,\ell/\tau$ for the solutions. The measured viscosities were found to be approximately independent of shear rate over the range of $V_x$ examined, so we report the results at the highest shear rate (i.e., largest $V_x$) for which the relative statistical uncertainty was smallest. This procedure was repeated ten times for the reported results.

\section{Results and Discussion}
\label{sec:results}
To systematically assess the effects of using skewed collision cells on the transport properties identified in Sec.~\ref{sec:methods}, we considered four boxes with the same, equal edge length $L$ but increasing amounts of deformation (Table \ref{tab:boxes}). We used $L = 20\,\ell$ unless otherwise noted. With the chosen tilt factors, the periodic images of all boxes generate the same cubic lattice, so we expect all four boxes should have similar finite-size effects. Indeed, numerical evaluation of Eq.~\eqref{eqn:diffusivity-ewald} gave the same value of $\xi$ in Eq.~\eqref{eqn:diffusivity-finite} for all four boxes and so the same finite-size correction to the self-diffusion coefficient. Since our primary interest is to compare effects of skewness between boxes, we will report all values as measured in the simulations to simplify the analysis.
\begin{table}
    \centering
    \caption{Simulation boxes used for simulations. All boxes had the same edge lengths, $L_x = L_y = L_z = L$, but different tilt factors.}
    \begin{tabular}{cccc}
        box & $f_{xy}$ & $f_{xz}$ & $f_{yz}$ \\ \hline
        A & 0 & 0 & 0 \\
        B & 1 & 0 & 0 \\
        C & 1 & 1 & 0 \\
        D & 1 & 1 & 1
    \end{tabular}
    \label{tab:boxes}
\end{table}

\subsection{Solvent}
\begin{figure}
    \centering
    \includegraphics{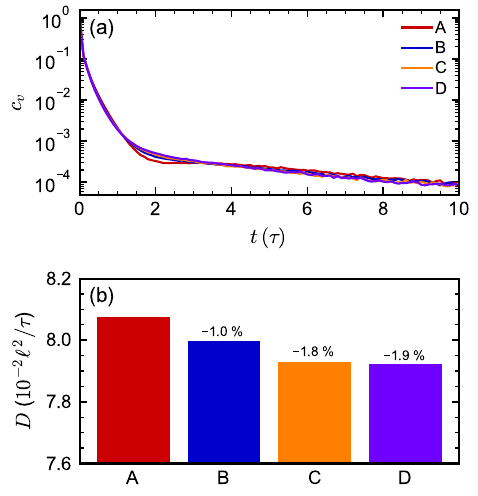}
    \caption{(a) Velocity autocorrelation function and (b) self-diffusion coefficient of the pure solvent with density $\rho = 5/\ell^3$ in Boxes A--D with $L=20\,\ell$. Percentages above bars in (b) indicate relative differences with respect to Box A.}
    \label{fig:solvent-vacf}
\end{figure}

We first examined the influence of skewing the collision cells on the dynamics of the pure solvent. The VACFs for the solvent in all four boxes exhibited the expected rapid short-time decay, followed by the long-time tail $t^{-3/2}$ [Fig.~\ref{fig:solvent-vacf}(a)] \cite{ripoll:physrev:2005, ernst:physrev:1971}. All VACFs were similar to each other; however, there was a small but noticeable deviation between them around $t = 2\,\tau$. The VACFs for Boxes B--D decreased less than that of Box A at this time, then somewhat faster at long times. We verified that qualitatively similar behavior was also observed at a larger solvent density ($\rho=10/\ell^3$, Fig.~S1). We also confirmed that this difference was not a trivial finite-size effect by repeating our calculations in a larger simulation box ($L = 100\,\ell$, Fig.~S2), finding essentially the same differences in the VACFs for Boxes A and B.

Differences in the VACF are expected to directly connect to differences in the self-diffusion coefficient $D$ because of the Green--Kubo relation between them.\cite{allen:book:2017} Indeed, $D$ determined from the MSD also showed small but consistent differences as the skewness of the collision cells increased [Fig.~\ref{fig:solvent-vacf}(b)]. It progressively decreased from Box A to Box D, with a largest deviation of $-1.9\%$ with respect to Box A observed in Box D. This decrease in $D$ is consistent with faster decay of the VACF in the skewed boxes.

\begin{figure}
    \centering
    \includegraphics{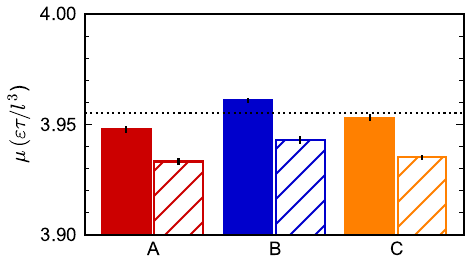}
    \caption{Shear viscosity of pure solvent with density $\rho = 5/\ell^3$ in Boxes A--C with $L = 20\,\ell$ obtained using RNES (filled bars) or force-driven simulations (hatched bars). The dotted line denotes the theoretically predicted viscosity for cubic collision cells.}
    \label{fig:solvent-viscosity}
\end{figure}

We additionally measured the shear viscosity $\mu$ of the pure solvent using both force-driven simulations and RNES. Only Boxes A--C were simulated due to implementation details of the RNES in HOOMD-blue. Previous simulations have confirmed that theoretical expressions for $\mu$, derived from kinetic theory for cubic collision cells, are highly accurate \cite{kikuchi:jcp:2003, tuzel:physrev:2003, ripoll:physrev:2005}. Consistent with these established results, the shear viscosity in Box A obtained using RNES was in close agreement with the theoretical prediction (Fig.~\ref{fig:solvent-viscosity}). Skewness in Boxes B and C led to nontrivial changes in $\mu$: it was larger in both Boxes B and C than in A, but it was less in the more skewed Box C than in Box B. The shear viscosity measured using the force-driven simulations was consistently somewhat smaller than from RNES but showed the same trends.

\subsection{Solutions}
The results in the previous section demonstrate that skewing the collision cells leads to small but statistically significant differences in the transport properties of the pure solvent. A direct practical implication of this behavior is that well-known analytical expressions for the solvent diffusivity and viscosity derived from kinetic theory for cubic collisions may be insufficiently accurate \cite{kikuchi:jcp:2003, ihle:physrev:2003a, ihle:physrev:2003b}, and these transport properties may need to be reanalyzed or simulated instead when using non-cubic collision cells. It is not immediately clear, though, whether skewness dependence of comparable magnitude should be expected for solutes embedded in an MPCD solvent. We hence next characterized the transport properties of solutions of nearly hard spheres. 

\begin{figure}
    \centering
    \includegraphics{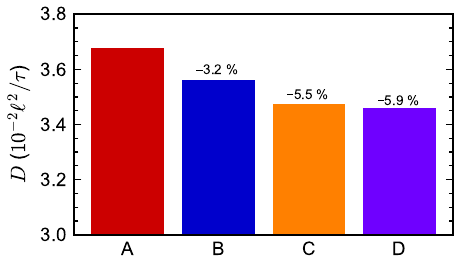}
    \caption{Self-diffusion coefficients of nearly hard spheres with density $\rho_{\rm s} = 0.2/\ell^3$ in solvent with density $\rho = 5/\ell^3$ in Boxes A--D with $L = 20\,\ell$. Percentages above bars indicate relative differences with respect to Box A.}
    \label{fig:solution-diffusion}
\end{figure}

Similarly to the pure solvent, the self-diffusion coefficient $D$ of the nearly-hard-sphere solutes progressively decreased from Box A to Box D as the skewness of the collision cells increased (Fig.~\ref{fig:solution-diffusion}). Box D exhibited the largest deviation of $-5.9\%$ with respect to Box A, which is approximately 3 times larger than that for the pure solvent and indicates the effects of skewness may be more sizable for the solutes. For comparison, we performed analogous isothermal--isochoric (NVT) simulations without solvent using either a Bussi thermostat \cite{bussi:jcp:2007} (time constant $1.0\,\tau$) or a Langevin thermostat (friction coefficient $1.0\,m/\tau$) \cite{allen:book:2017}. The values of $D$ computed in Boxes A and B were statistically indistinguishable for both thermostats (Fig.~S3 and Table S2), confirming that the effects of skewness on the solute self-diffusion constant are associated with the MPCD solvent.

\begin{figure}
    \centering
    \includegraphics{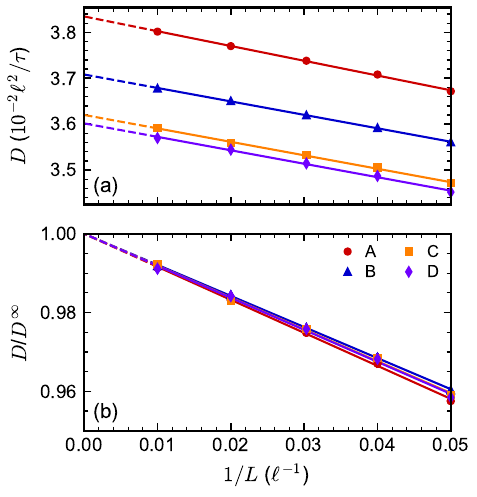}
    \caption{(a) Self-diffusion coefficients of nearly hard spheres with density $\rho_{\rm s} = 0.2/\ell^3$ in solvent with density $\rho = 5/\ell^3$ in Boxes A--D with $L = \{20, 25, 33, 50, 100\}\,\ell$. Solid symbols are simulated values, solid lines are linear fits to the simulated values, and dashed lines are linear extrapolation. (b) The same data as in (a) but with $D$ normalized by the self-diffusion coefficient extrapolated to infinite box size $D^{\infty}$. Uncertainties in the simulated data are approximately the size of or smaller than the symbols.}
    \label{fig:solution-finite}
\end{figure}

Given the more substantial dependence of $D$ on skewness for the solutes, we next checked whether this behavior might be due to finite-size effects. We expected that it should not based on our evaluation of Eq.~\eqref{eqn:diffusivity-ewald}, but this analytical correction has not been tested previously for MPCD with skewed simulation boxes. We accordingly conducted a series of simulations in boxes with the same tilt factors but $L$ ranging from $20\,\ell$ to $100\,\ell$. We decreased the number of independent replicas that we simulated as the box size increased; see Table S1 for a complete listing. We expected $D$ to have a linear dependence on $1/L$ based on Eq.~\eqref{eqn:diffusivity-finite} with the same slope for all boxes if the solution viscosity does not depend significantly on skewness. We confirmed this relationship in our simulations [Fig.~\ref{fig:solution-finite}(a)], but there were similar differences in $D$ with skewness even after extrapolating to infinite box size using linear regression. Interestingly, though, we noted that our regression did not give the same slope for the four boxes [Fig.~\ref{fig:solution-finite}(b)]. The points for the smallest boxes might deviate somewhat from the linear fit, but discarding them did not qualitatively change our findings. Additionally, analogous simulations performed for Boxes A and B using a Bussi thermostat yielded identical slopes, confirming that the spurious skewness dependence is an artifact of the MPCD algorithm (Fig.~S4).

\begin{figure}
    \centering
    \includegraphics{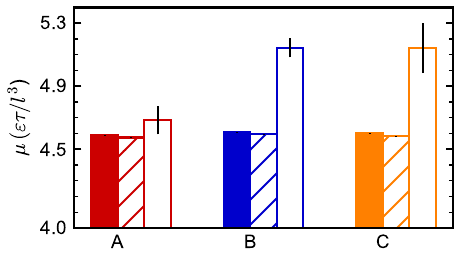}
    \caption{Shear viscosity of nearly hard spheres with density $\rho_{\rm{s}} = 0.2/\ell^3$ in solvent with density $\rho = 5/\ell^3$ in Boxes A--C with $L = 20\,\ell$ using RNES (filled bars), force-driven simulations (hatched bars), or extracted from the slope of fits in Fig.~\ref{fig:solution-finite} using Eq.~\eqref{eqn:diffusivity-finite} (open bars).} 
    \label{fig:solution-viscosity}
\end{figure}

One possible explanation for the differences between the slopes of the fits in Fig.~\ref{fig:solution-finite} is a dependence of solution viscosity on collision-cell skewness. We extracted a nominal solution viscosity for Boxes A--C from these slopes and compared it to that computed directly using both force-driven simulations and RNES (Fig.~\ref{fig:solution-viscosity}). The viscosity computed directly from the nonequilibrium simulations had only a weak dependence on skewness, as for the solvent (Fig.~\ref{fig:solvent-viscosity}). For Box A, the nominal solution viscosity from finite-size scaling agreed reasonably well with the values computed directly. However, for Boxes B and C, the nominal viscosity from finite-size scaling was significantly larger than values obtained directly. The inconsistency in viscosity extracted from finite-size scaling with that measured directly indicates a different underlying cause for the changes in the solute self-diffusion coefficient with skewness.

\begin{figure}
    \centering
    \includegraphics{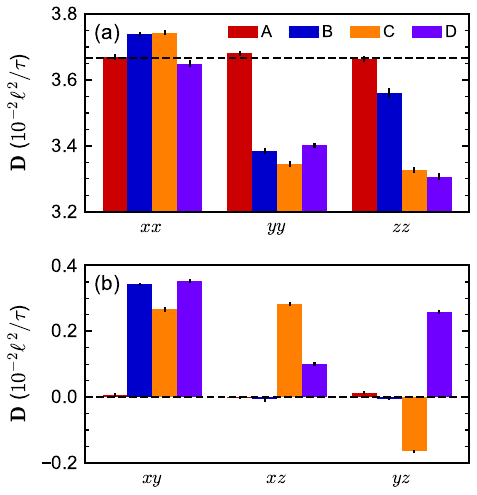}
    \caption{Self-diffusion tensor (a) diagonal and (b) unique off-diagonal components of nearly hard spheres with density $\rho_{\rm s} = 0.2/\ell^3$ in solvent with density $\rho = 5/\ell^3$ in Boxes A--D with $L = 20\,\ell$. The dashed lines indicate the expectation for isotropic diffusion based on $D$ for Box A.}
    \label{fig:solution-tensor}
\end{figure}

We hypothesized that a possible cause could be unphysical anisotropy in the solute diffusion that is introduced by the skewed collision cells. We computed the diffusion tensor $\vv{D}$ for the solutes in Boxes A--D (Fig.~\ref{fig:solution-tensor}) using the same displacement data as was used to determine $D$. For Box A, the diagonal components of the diffusion tensor were similar to the scalar self-diffusion coefficient $D$ and the off-diagonal components were close to zero; the cubic collision cells of Box A produced essentially isotropic solute diffusion, as expected. For Boxes B--D, though, the diagonal components of the diffusion tensor deviated both from $D$ for Box A and from each other, and the off-diagonal elements were nonzero; the solute diffusion became strongly anisotropic. Hence, even though the scalar diffusivity $D$, which is an isotropic average of $\vv{D}$, only changed by a few percent in Boxes B--D, skewing the collision cells introduced strong directional biases that qualitatively and spuriously alter the solute dynamics.

\begin{figure}
    \centering
    \includegraphics{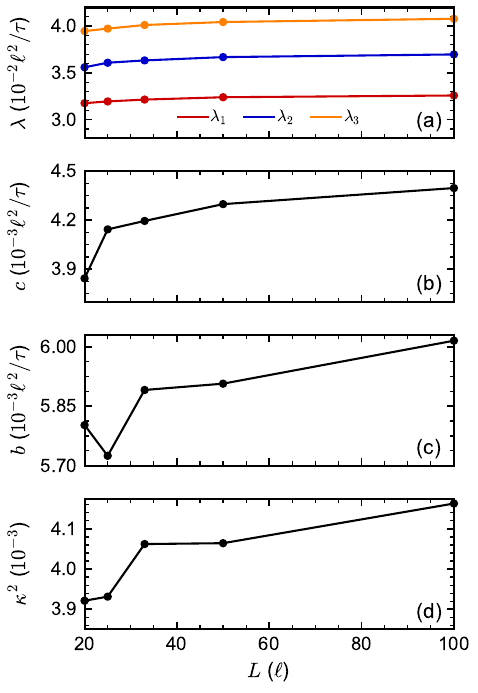}
    \caption{(a) Eigenvalues, (b) acylindricity, (c) asphericity, and (d) relative shape anisotropy for self-diffusion tensor of nearly hard spheres with density $\rho_{\rm s} = 0.2/\ell^3$ in solvent with density $\rho = 5/\ell^3$ in Box B with varying $L$.}
    \label{fig:solution-anisotropy}
\end{figure}

We characterized the anisotropy of the diffusion tensor in Box B by computing the acylindricity $c$, asphericity $b$, and relative shape anisotropy $\kappa^2$,
\begin{align}
c &= \lambda_2 - \lambda_3 \\
b &= \lambda_1 - \frac{1}{2}(\lambda_2 + \lambda_3) \\
\kappa^2 &= \frac{3}{2}\frac{\lambda_1^2 + \lambda_2^2 + \lambda_3^2}{(\lambda_1 + \lambda_2 + \lambda_3)^2} - \frac{1}{2},
\end{align}
from the eigenvalues $\lambda_1 \geq \lambda_2 \geq \lambda_3$ of $\vv{D}$. All measures confirmed anisotropy in Box B with $L=20\,\ell$ (Fig.~\ref{fig:solution-anisotropy}). We then repeated this analysis for simulation boxes of increasing size to test whether the anisotropy was a finite-size effect that might decrease with increasing box size. We found the opposite, with anisotropy actually increasing with $L$. Similar analysis for the cubic Box A showed much smaller amounts of anisotropy in the smallest boxes that vanished vanished as the box size increased, as expected for isotropic diffusion (Fig.~S5). The anisotropy of $\vv{D}$ in the skewed Boxes B--D provides a plausible explanation for the larger-than-expected values of $\mu$ extracted from the finite-size scaling of $D$: anisotropic diffusion is unphysical for a bulk solution and so inconsistent with the analysis that leads to Eq.~\eqref{eqn:diffusivity-finite}.

Lastly, we investigated the generality of our findings by measuring the solute self-diffusion coefficient in Boxes A and B for varying solute densities $\{0.2, 0.4, 0.6\}/\ell^3$ and solvent  densities $\{5,10,25\}/\ell^3$. The solute self-diffusion coefficient in Box B was consistently smaller than in Box A, with deviations ranging from $-1.8\%$ to $-7.0\%$ (Table \ref{tab:solution-density}). These results indicate that collision-cell skewness is likely to produce dynamical artifacts for solutes across a wide range of conditions relevant to many practical uses of MPCD simulations.
\begin{table}
    \centering
    \caption{Percent difference in self-diffusion coefficient of nearly hard spheres in Box B relative to in Box A, both with $L = 20\,\ell$, for varied solvent and solute densities.}
    \begin{tabular}{c|ccc}
        solute density & \multicolumn{3}{c}{solvent density $\rho$ ($\ell^{-3}$)} \\
        $\rho_{\rm s}$ ($\ell^{-3}$) & 5 & 10 & 25 \\ \hline
        0.2 & $-3.0 \pm 0.2$ & $-2.3 \pm 0.3$ & $-1.8 \pm 0.2$ \\
        0.4 & $-4.7 \pm 0.4$ & $-4.5 \pm 0.1$ & $-4.0 \pm 0.2$ \\
        0.6 & $-6.7 \pm 0.4$ & $-6.7 \pm 0.1$ & $-7.0 \pm 0.5$ \\
    \end{tabular}
    \label{tab:solution-density}
\end{table}

\section{Conclusions}
\label{sec:conclusions}
We modified the standard MPCD algorithm to use skewed collision cells that are arranged in a grid aligned with a parallelepiped simulation box, and we investigated how skewness affects transport properties of both the pure solvent and solutions of nearly hard spheres. Skewness was found to have a small but statistically significant impact on the velocity autocorrelation function, self-diffusion coefficient, and shear viscosity of the pure solvent. For example, the self-diffusion coefficient of the solvent decreased by up to 1.9\% with skewed cells compared to standard cubic cells. Qualitatively similar behavior was also found for the self-diffusion coefficient of the nearly-hard-sphere solutes, but the dependence on skewness was stronger than for the pure solvent. After performing a finite-size scaling analysis of the solute self-diffusion coefficient and making independent measurements of the solution shear viscosity, we identified anisotropy in the solute diffusion as a probable cause of the skewness dependence. This anisotropy is unphysical and not present when using cubic collision cells in MPCD, so we regard it as an artifact introduced by using skewed collision cells.

Dependence of transport properties on collision-cell skewness is undesirable because the cells are an artificial but necessary construction in the MPCD algorithm, and ideally, the physics in the simulation would be independent of cell shape. As such, skewed collision cells should be used in MPCD simulations with caution. We suspect that a grid-free scheme to bin particles into collision cells, such as that proposed in Ref.~\citenum{muhlbauer:physrev:2017}, may help mitigate dependence on skewness and warrants further investigation.

\section*{Supplementary material}
See the supplementary material for additional simulation details and characterizations of both pure solvent and solutions.

\section*{Conflicts of interest}
The authors have no conflicts to disclose.

\section*{Data availability}
The data that support the findings of this study are available from the authors upon reasonable request.

\section*{Acknowledgments}
This material is based upon work supported by the National Science Foundation under Award Nos.~2310724 and 2310725. J.C.P. gratefully acknowledges partial support from the Welch Foundation (Grants E-1882 and V-E-0001). This work was completed with resources provided by the Auburn University Easley Cluster and the Hewlett Packard Enterprise Data Science Institute at the University of Houston.

\bibliography{references}

\end{document}


\title{Supplementary material for ``Effects of skewing collision cells on transport properties in multiparticle collision dynamics simulations''}

\author{Jinny Cha}
\affiliation{Department of Chemical Engineering, Auburn University, Auburn, AL 36849, USA}

\author{Wilfred Kwabena Darko}
\affiliation{Department of Chemical and Biomolecular Engineering, University of Houston, Houston, TX 77204, USA}

\author{Jeremy C. Palmer}
\affiliation{Department of Chemical and Biomolecular Engineering, University of Houston, Houston, TX 77204, USA}

\author{Michael P. Howard}
\email{mphoward@auburn.edu}
\affiliation{Department of Chemical Engineering, Auburn University, Auburn, AL 36849, USA}

\maketitle

\begin{table}[!h]
    \centering
    \caption{Box sizes $L$ and number of replicas used for solute self-diffusion simulations.}
    \begin{tabular}{c|c}
        $L$ & number of replicas \\ \hline
        20 & 10 \\
        25 & 10 \\
        33 & 10 \\
        50 & 5 \\
        100 & 5 \\
    \end{tabular}
\end{table}

\begin{figure}[!h]
    \centering
    \includegraphics{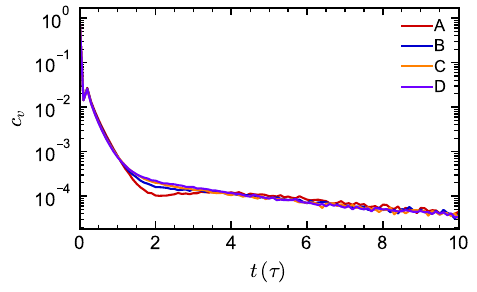}
    \caption{Same as Fig.~2(a) but for density $\rho = 10/\ell^3$.}
\end{figure}

\begin{figure}[!h]
    \centering
    \includegraphics{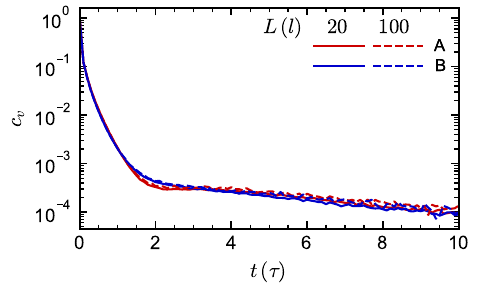}
    \caption{Velocity autocorrelation function of the pure solvent with density $\rho = 5/\ell^3$ in Boxes A and B with $L=\{20,100\}\,\ell$.}
    \label{fig:solvent-vacf-finite}
\end{figure}

\begin{figure}[!h]
    \centering
    \includegraphics{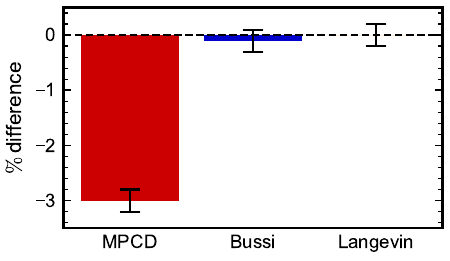}
    \caption{Percent difference in self-diffusion coefficient of nearly hard spheres in Box B relative to in Box A, both with $L = 20\,\ell$, simulated using MPCD at $\rho=5/\ell^3$ or with NVT MD using either a Bussi or Langevin thermostat.}
    \label{fig:solution-methods}
\end{figure}

\begin{table}[!h]
    \centering
    \caption{Values of self-diffusion coefficient (in $\ell^2/\tau$) and percent difference in Box B relative to Box A using the same methods as in Fig.~\ref{fig:solution-methods} and $L = \{20,100\}\ell$. Uncertainty in the last digit is given in parentheses. The values of $D$ are not expected to agree between methods.}
    \begin{tabular}{c|ccc|ccc}
        & \multicolumn{3}{c|}{$L = 20\,\ell$} & \multicolumn{3}{c}{$L = 100\,\ell$} \\ \cline{2-7}
        & $D$ in A & $D$ in B & \% difference & $D$ in A & $D$ in B & \% difference \\ \hline
        MPCD & 0.03672(7) & 0.03561(4) & -3.0(2) & 0.03801(1) & 0.03677(2) & -3.27(6) \\
        NVT & 0.4170(5) & 0.4167(5) & -0.1(2) & 0.42414(7) & 0.42401(7) & -0.03(2) \\
        Langevin & 0.2828(4) & 0.2827(3) & 0.0(2) & 0.28276(4) & 0.28260(4) & -0.06(2) \\
    \end{tabular}
    \label{tab:solution-percent-difference}
\end{table}

\begin{figure}[!h]
    \centering
    \includegraphics{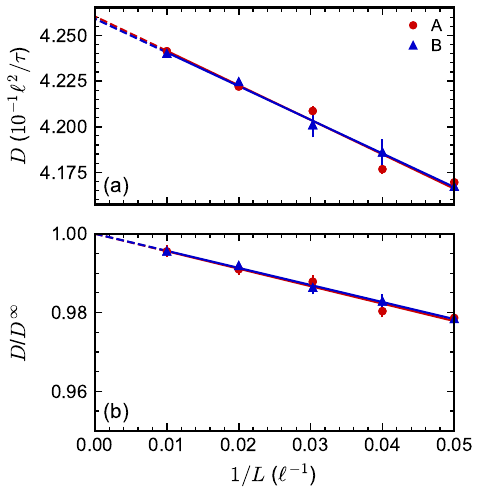}
    \caption{Same as Fig.~5 but using NVT MD with a Bussi thermostat and only for Boxes A and B.}
    \label{fig:solution-nvt}
\end{figure}

\begin{figure}[!h]
    \centering
    \includegraphics{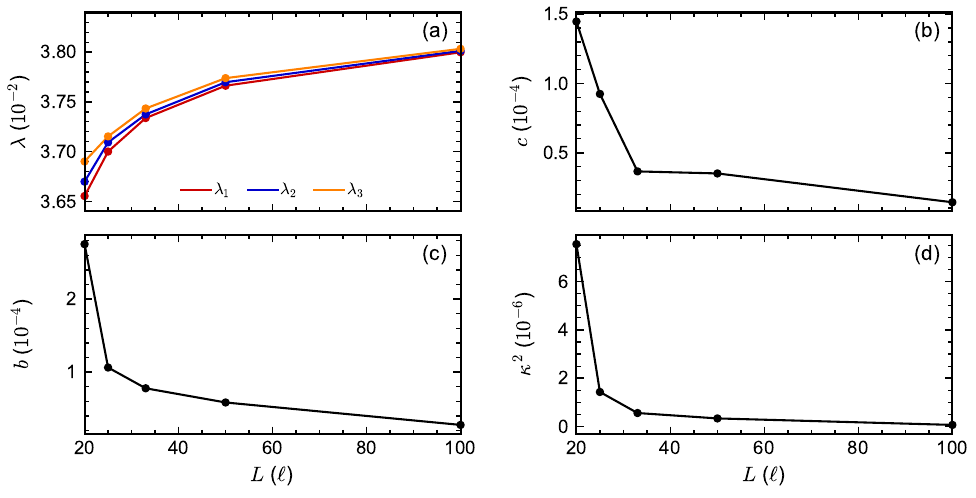}
    \caption{Same as Fig.~8 but for Box A.}
    \label{fig:solution-anisotropy-boxA}
\end{figure}

\clearpage
\begin{table}[!h]
    \caption{Measured average rate of momentum transfer $\dot p_x$ (in $m\ell/\tau$), shear rate $\dot\gamma$ (in $1/\tau$) and calculated viscosity $\mu$ (in $\varepsilon \tau/\ell^3$) for a pure solvent with density $\rho = 5/\ell^3$ in Boxes A--C using RNES. Results are given for different target velocities $V_x$ (in $\ell/\tau$). Uncertainties in the last digit are given in parentheses.}
    \label{tab:rnes-data-solvent}
    \centering
    \begin{tabular}{c|ccc|ccc|ccc}
         & \multicolumn{3}{c|}{A} & \multicolumn{3}{c|}{B} & \multicolumn{3}{c}{C} \\ \cline{1-10}
         $V_x$ & $\dot p_x$ & $\dot\gamma$ & $\mu$ & $\dot p_x$ & $\dot\gamma$ & $\mu$ & $\dot p_x$ & $\dot\gamma$ & $\mu$ \\ \hline
         1 & 39.99754(8) & 0.01267(1) & 3.945(3) & 39.99778(1) & 0.01264(2) & 3.957(5) & 39.99778(1) & 0.01264(2) & 3.957(5) \\
         5 & 108.4(2) & 0.03433(6) & 3.948(2) & 108.9422(9) & 0.03438(1) & 3.961(1) & 108.9334(8) & 0.03445(1) & 3.953(2) \\
    \end{tabular}
\end{table}

\begin{table}[!h]
    \caption{Same as Table \ref{tab:rnes-data-solvent} but for nearly hard spheres with density $\rho_{\rm s} = 0.2/\ell^3$ in solvent with density $\rho = 5/\ell^3$ .}
    \label{tab:rnes-data-solution}
    \centering
    \begin{tabular}{c|ccc|ccc|ccc}
         & \multicolumn{3}{c|}{A} & \multicolumn{3}{c|}{B} & \multicolumn{3}{c}{C} \\ \cline{1-10}
         $V_x$ & $\dot p_x$ & $\dot\gamma$ & $\mu$ & $\dot p_x$ & $\dot\gamma$ & $\mu$ & $\dot p_x$ & $\dot\gamma$ & $\mu$ \\ \hline
         0.1 & 6.66653(1) & 0.00182(1) & 4.58(3) & 6.66654(2) & 0.00180(1) & 4.63(2) & 6.66655(2) & 0.00181(1) & 4.61(2) \\
         0.5 & 33.33241(2) & 0.00907(1) & 4.595(5) & 33.33253(2) & 0.00903(1) & 4.616(5) & 33.33248(2) & 0.00906(1) & 4.600(5) \\
         1 & 66.6623(1) & 0.01815(1) & 4.591(3) & 66.66269(2) & 0.01808(1) & 4.610(2) & 66.66270(2) & 0.01813(1) & 4.597(2) \\
         5 & 176.3(3) & 0.04800(7) & 4.5920(5) & 177.224(2) & 0.04807(1) & 4.6088(9) & 177.209(2) & 0.04814(1) & 4.6011(6) \\
    \end{tabular}
\end{table}